\documentclass[11pt]{article}

\usepackage[dvips]{graphicx}
\usepackage{amssymb}
\usepackage{alltt}
\usepackage{afterpage}

\newcommand{\grad}{\nabla}

\newcommand{\half}{{\textstyle{\frac{1}{2}}}}

\newcommand{\Ref}[1]{\mbox{\rm{(\ref{#1})}}}

\newcommand{\R}{\mbox{${\mathbb R}$}}

\usepackage{art11mod}
\usepackage{numinsec}
\usepackage{float}
\usepackage{url}

\newcommand{\cD} {\mbox{$\cal D$}}

\newcommand{\cS} {\mbox{$\cal S$}}

\newcommand{\ctimeg}{\mbox{$ \Omega_{\scriptscriptstyle T,G} $}}
\newcommand{\cmemg}{\mbox{$ \Omega_{\scriptscriptstyle M,G} $}}
\newcommand{\ctimeh}{\mbox{$ \Omega_{\scriptscriptstyle T,H} $}}
\newcommand{\cmemh}{\mbox{$ \Omega_{\scriptscriptstyle M,H} $}}
\newcommand{\fext}{\mbox{$ f_{\scriptscriptstyle E} $}}
\newcommand{\rhom}[1]{\mbox{$ \rho_{\scriptscriptstyle M}#1 $}}

\newcommand{\cops} {\mbox{\sf \footnotesize COPS}}

\newcommand{\tron} {\mbox{\sf \footnotesize TRON}}

\newcommand{\fsqp} {\mbox{\sf \footnotesize FSQP}}
\newcommand{\lancelot} {\mbox{\sf \footnotesize LANCELOT}}
\newcommand{\loqo} {\mbox{\sf \footnotesize LOQO}}

\newcommand{\minpack} {\mbox{\sf \footnotesize MINPACK-2}}

\renewcommand{\R}{\mbox{${\mathbb R}$}}

\newfloat{Algorithm}{thp}{loa}[section]

\begin{document}
\setlength{\baselineskip}{14pt}

\pagestyle {empty}

\vspace{1.75in}

\begin{center}

ARGONNE NATIONAL LABORATORY

9700 South Cass Avenue

Argonne, Illinois  60439

\vspace{1.5in}

{\large
{\bf 
AUTOMATIC DIFFERENTIATION TOOLS IN OPTIMIZATION SOFTWARE 
}
}

\vspace{.5in}

{\bf Jorge J. Mor\'e}

\vspace{.5in}

Mathematics and Computer Science Division

\vspace{.25in}

Preprint ANL/MCS-P859-1100

\vspace{.5in}

November 2000
\end{center}

\vspace{2.0in}

\bigskip

\par\noindent
This work was supported by the Mathematical, Information, and
Computational Sciences Division subprogram of the Office of Advanced
Scientific Computing, U.S. Department of Energy, under Contract
W-31-109-Eng-38, and
by the the National Science Foundation 
(Information Technology Research) grant CCR-0082807. 

\newpage
\mbox{}
\newpage

\pagestyle {plain}
\setcounter {page}{1}

\begin {center}

{\large\bf Automatic Differentiation Tools in Optimization Software}

\bigskip

{\bf Jorge J. Mor\'e}

\end {center}

\begin{abstract}
We discuss the role of automatic
differentiation tools in optimization software.
We emphasize issues that are important to large-scale optimization
and that have proved useful in the installation of
nonlinear solvers in the NEOS Server.
Our discussion centers on the computation of the
gradient and Hessian matrix for partially separable functions
and shows that the gradient and Hessian matrix
can be computed with guaranteed bounds in 
time and memory requirements.
\end{abstract}

\footnotetext
{
Mathematics and Computer Science Division,
Argonne National Laboratory,
9700 South Cass Avenue,
Argonne, Illinois 60439
({\tt more@mcs.anl.gov}).
This work was supported by the Mathematical, Information, and
Computational Sciences Division subprogram of the Office of Advanced
Scientific Computing, U.S. Department of Energy, under Contract
W-31-109-Eng-38, and
by the the National Science Foundation 
(Information Technology Research) grant CCR-0082807. 
}

\section {Introduction}

\label{intro}

Despite advances in automatic differentiation algorithms and software,
researchers disagree on the value of incorporating
automatic differentiation tools in optimization software.
There are various reasons for this state of affairs.
An important reason seems to be that little
published experience exists on the effect of automatic differentiation
tools on realistic problems, and thus users worry that automatic
differentiations tools are not applicable to their problems or 
are too expensive
in terms of time or memory.
Whatever the reasons, few optimization codes 
incorporate automatic differentiation tools.

Without question, incorporating automatic differentiation
tools into optimization is not only useful but, in many cases,
essential in order to promote the widespread use of
state-of-the-art optimization software.
For example, a Newton method for the
solution of large bound-constrained problems
\[
\min \left \{ f(x) : x_l \le x \le x_u \right \},
\]
where $ f : \R^n \mapsto \R $ and $ x_l $ and $ x_u $ define
the bounds on the variables, requires that the user provide
procedures for evaluating the function $ f(x) $ and also
the gradient $ \grad f (x) $,
the sparsity pattern of the Hessian matrix $ \grad ^2 f (x) $,
and the  Hessian matrix $ \grad ^2 f (x) $.
The demands on the user increase for the
constrained optimization problem
\[
\min \left \{ f(x) : x_l \le x \le x_u , \ c_l \le c(x) \le c_u \right \} ,
\]
where $ c : \R^n \mapsto \R^m $ are the nonlinear constraints.
In this case the user must also provide the sparsity 
pattern and the Jacobian matrix $ c'(x) $
of the constraints. In some cases the user may even be asked
to provide the Hessian matrix of the Lagrangian
\begin{equation}
\label{intro:lagrangian}
L (x,u) = f(x) + \langle u , c(x) \rangle
\end{equation}
of the optimization problem.
The time and effort required to obtain this information and
verify their correctness can be large even for simple
problems. Clearly, any help in simplifying this effort
would promote the use of the software.

In spite of the advantages offered by automatic
differentiation tools, relatively little effort
has been made to interface optimization software with automatic
differentiation tools.
Dixon \cite{Dixon1991a,Dixon1991b} was an early proponent
of the integration of automatic differentiation with optimization,
but to our knowledge
Liu and Tits \cite{LiuTits1997} were the
first to provide interfaces between a general nonlinear
constrained optimization solver (\fsqp) and automatic
differentiation tools (ADIFOR).

Modeling languages for optimization (for example,
AMPL~\cite{ampl-home} and
GAMS~\cite{gams-home}) provide environments for solving
optimization problems that deserve emulation. 
These environments package the ability to calculate derivatives,
together with state-of-the-art optimization solvers and
a language that facilitates modeling, to yield an extremely
attractive problem-solving environment.

The NEOS Server for Optimization \cite{neos-home} is another
problem-solving environment that integrates automatic differentiation
tools and state-of-the-art optimization solvers.  Users choose a solver
and submit problems via the Web, email (\url{neos@mcs.anl.gov}), or a
Java-enabled submission tool. When a submission arrives, NEOS parses
the submission data and relays that data to a computer associated with
the solver. Once results are obtained, they are sent to NEOS, which
returns the results to the user. Submissions specified in Fortran
are processed by ADIFOR \cite{ADIFOR2,adifor2-revC}, 
while C submissions are handled
by ADOL-C \cite{ADOL-C}.
Since the initial release in 1995, the NEOS Server has continued
to add nonlinear optimization solvers with an emphasis on
large-scale problems, and the current version contains
more than a dozen different nonlinear optimization solvers.

Users of a typical computing
environment would like to solve optimization problems
while only requiring that the user provide a specification
of the problem; all other quantities required by the software (for
example, gradients, Hessians, and sparsity patterns) would be generated automatically.
Optimization modeling languages and the NEOS Server provide this
ability, but as noted above, users of nonlinear optimization solvers
are usually asked to provide derivative information.

Our goal in this paper is to discuss techniques for using automatic
differentiation tools in large-scale optimization software.
We highlight issues that are relevant to solvers in the 
NEOS Server. For recent work on the interface between automatic differentiation
tools and large-scale solvers, see 
\cite{AD2K-JA00,AD2K-PDH00}.
We pay particular attention to the computation of 
second-order (Hessian) information since there is evidence
that the use of second-order information is crucial
to the solution of large-scale problems. The main concern is
the cost of obtaining second-order information.
See \cite{JA97,RG00,AG00} for related work.

We note that at present most optimization software for large-scale
problems use only first-order derivatives.
Indeed, of the nonlinear solvers available in the
NEOS Server, only \lancelot, \loqo, and \tron\  accept second-order
information. We expect this situation to change, however, as automatic
differentiation tools improve and provide 
second-order information with the same reliability and
efficiency as are currently available for first-order information.

\section{Partially Separable Functions}

We consider the computation of the gradient and Hessian
matrix of a partially
separable function, that is, a function 
$ f : \R^n \mapsto \R $ of the form
\begin{equation}
\label{psf:decomposition}
f ( x) = \sum_{k=1}^m f_k(x) ,
\end{equation}
where the component functions $ f_k : \R^n \mapsto \R $ are
such that the \textit{extended} function
\[
\fext (x ) =
\left ( 
\begin{array}{c}
f_1 (x) \\
\vdots \\
f_m(x)
\end{array}
\right )
\]
has a sparse Jacobian matrix. Our techniques are geared to
the solution of large-scale optimization problems.
For an extensive treatment of techniques for computing
derivatives of general and partially separable functions
with automatic differentiation tools,
we recommend the recent book by Griewank \cite{AG00}.

Partially separable
functions were introduced by Griewank and Toint \cite{AG82a}.
They showed, in particular, that
$ f : \R^n \mapsto \R $ is partially
separable if and only if the Hessian matrix $ \grad ^ 2 f (x) $ is sparse.
Partially separable functions also arise
in systems of nonlinear equations and nonlinear least
squares problems. For example, if each component of the mapping
$ r : \R^n \mapsto \R^m $ is partially separable, then
\[
f (x) = \half \| r(x) \| ^ 2 
\]
is also partially separable. As another example, consider the
constrained optimization problem
\[
\min \left \{ f(x) : x_l \le x \le x_u , \ c_l \le c(x) \le c_u \right \} ,
\]
where
$ c : \R^n \mapsto \R^m $ specifies the constraints.
For this problem, the Lagrangian function $ L ( \cdot, u) $ defined
by \Ref{intro:lagrangian}
is partially separable if $f$ and all the components of
the mapping $c$ are partially separable.
For specific examples note that the functions
$f$ and $c$ in the parameter estimation
and optimal control optimization problems in the \cops\ \cite{EDD00}
collection are partially separable.

We are interested in computing
the gradient and the Hessian of a partially
separable function with guaranteed bounds in terms of both computing
time and memory requirements. We require that the computing time
be bounded by a multiple of the computing time of the function,
that is,
\begin{equation}
\label{psf:ctime}
T \{ \grad f (x) \}  \le  \ctimeg \, T \{ f (x) \} , \qquad
T \{ \grad ^ 2 f (x) \}  \le  \ctimeh \, T \{ f (x) \} ,
\end{equation}
for constants
$ \ctimeg $ and $ \ctimeh $,
where $ T \{ \cdot \} $ is computing time.
We also require that
\begin{equation}
\label{psf:cmem}
M \{ \grad f (x) \}  \le  \cmemg \, M \{ f (x) \} , \qquad
M \{ \grad ^2 f (x) \}  \le  \cmemh \, M \{ f (x) \}
\end{equation}
for constants
$ \cmemg $ and $ \cmemh $, where
$ M \{ \cdot \} $ is memory. 

These are important requirements for large-scale problems.
In particular, if the constants in these expressions
are small and independent of the structure of the 
extended function $ \fext $,
then the computational requirements of an iteration of Newton's
method are comparable with those of a limited-memory Newton's
method.

The constants in \Ref{psf:ctime} and \Ref{psf:cmem} 
can be bounded in terms of a measure of the sparsity of the
extended function. We use $ \rhom{} $, where
\[
\rhom{} \equiv \max \{ \rho_i \} ,
\]
and $ \rho_i $ is the number of nonzeros in the 
$i$th row of $ \fext '(x) $. We can also view $ \rhom{} $ as the
largest number of variables in any of the component functions.

Decompositions \Ref{psf:decomposition}
with the number $m$
of element functions of order $n$, 
and with $ \rhom{} $ small and independent of $n$,
are preferred.
Since the number of nonzeros in
the Hessian $ \grad^2 f (x) $ is no more than $ m \rhom{} $,
decompositions with these properties
are guaranteed to have sparse Hessian matrices.
Discretizations of parameter estimation and optimal control problems,
for example, have these properties because in these problems each element
function represents the contributions from an interval or an element
in the discretization.

One of the aims of this paper is to present numerical evidence 
that we can compute the
gradient $ \grad f (x) $ and the Hessian matrix 
 $ \grad ^2 f (x) $ of a partially
separable function with 
\begin{equation}
\label{psf:bound}
\ctimeg \le \kappa_1 \rhom{} , \qquad
\ctimeh \le \kappa_2 \rhom{^2} ,
\end{equation}
where $ \kappa_1 $ and $ \kappa_2 $ are constants of modest size
and independent of $ \fext $. We normalize
$ \ctimeg $ by $ \rhom{} $ because the techniques in Section \ref{sec:gradients}
require at least $ \rhom{} $ functions evaluations to
estimate the gradient. 
Similarly, the number of gradient evaluations needed to estimate
the Hessian matrix by the techniques in Section \ref{sec:hessians}
is at least $ \rhom{} $. Thus, these techniques require
at least $ \rhom{^2} $ function 
evaluations to estimate the Hessian matrix.

\section{Computing Gradients}

\label{sec:gradients}

We now outline the techniques that we use for computing the
gradients of partially separable functions.
For additional information on
the techniques in this section, see
\cite{BOUARICHA94,B95fa}.

Computing the gradient of a partially separable function
so that the bounds \Ref{psf:ctime} and \Ref{psf:cmem} 
are satisfied is based on the observation, due to
Andreas Griewank, that
if $ f : \R^n \to \R $ is partially separable, then
\[
f(x) = \fext (x)^T e ,
\]
where $ e \in \R^m $ is the vector of all ones, and hence 
\begin{equation}
\label{grads:psf}
\grad f (x) = \fext '(x)^T e .
\end{equation}
We can then compute the gradient by computing the 
Jacobian matrix $ \fext ' (x) $.

At first sight the approach based on \Ref{grads:psf} does not look
promising, since we need to compute a Jacobian matrix and then
obtain the gradient from a matrix-vector product.  However, the key
observation is that the Jacobian matrix is sparse, while the
gradient is dense. Thus, we can use sparse techniques for
the computation of the extended Jacobian.


We could also use the reverse approach 
of automatic differentiation
to compute the gradient of $f$. The reverse approach works
directly on $f$ and does not require the partial separability
structure of $f$. Moreover, for the reverse approach,
\Ref{psf:ctime} holds with $ \ctimeg $ small and independent
of $ \rhom{} $. Theoretically $ \ctimeg \le 5 $, but practical
implementations may not satisfy this bound.
However, the memory requirements of the reverse approach
depend on the number of floating point operations needed
to compute $f$, and thus \Ref{psf:cmem} can be violated.
A careful comparison between the reverse approach and the
techniques described below would be of interest.

In this section we consider two
methods for computing the gradient of a partially separable function
via \Ref{grads:psf}. In the \textit{compressed} AD approach, automatic
differentiation tools are used to compute a
compressed form of the Jacobian matrix of the extended function
$ \fext $, while in 
the \textit{sparse} AD approach, automatic differentiation tools
are used to compute a sparse representation of the Jacobian
matrix  of the extended function.

In the compressed AD approach we assume that the sparsity
pattern of the Jacobian matrix $ \fext '(x) $ is known.
Given the sparsity pattern, we
partition the columns of the Jacobian matrix into groups of 
\textit{structurally orthogonal} columns, that is, 
columns that do not have a nonzero in
the same row position.
Given a partitioning of the columns into $p$
groups of structurally orthogonal columns, 
we determine  the Jacobian matrix by
computing the \textit{compressed Jacobian} matrix $ \fext '(x) V $,
where $ V \in \R ^ {n \times p } $.
There is a column of $V$ for each group, and
$ v_{i,j} \neq 0 $ only if the \textit{i}th column of
$ \fext '(x) $ is in the
\textit{j}th group.
Software for this partitioning problem \cite{TFC84} defines the
groups with an array
\texttt{ngrp} that sets the group for each column.

The extended Jacobian can be determined from the compressed
Jacobian matrix  $ \fext '(x) V $ by noting
that if column $j$ is in group $k$, then
\[
\langle e_i , \fext ' (x) V e_k \rangle = v_{i,j} 
\partial _{i,j} \fext (x) .
\]
Thus $ \partial _{i,j} \fext (x) $ can be recovered directly
from the compressed Jacobian matrix.

We note that for many sparsity patterns, the number
of groups $p$ needed to determine $ A \in \R^{m \times n} $ 
with a partitioning of the columns is small and independent of $n$.
In all cases there is a lower bound of $ p \ge \rhom{} $.
We also know \cite{TFC83} that if a matrix $ A $
can be permuted to a matrix
with bandwidth $ \textit{band} (A) $, then $ p \le \textit{band}(A) $.

The sparse AD approach uses a sparse data representation,
usually in conjunction with dynamic memory allocation, to carry out
all intermediate derivative computations.
At present, the 
SparsLinC library in ADIFOR~\cite{adifor2-revC}
is the only automatic differentiation tool with this capability.
The main advantage of the sparse AD approach over 
the compressed AD approach 
is that no knowledge of the sparsity pattern is required.
On the other hand, the sparse AD approach is almost always
slower, and can be significantly slower on vector machines.

In an optimization setting, a hybrid approach \cite{AB95} is the best
approach. With this strategy, the sparse AD approach is used to
obtain the sparsity pattern of the Jacobian matrix 
of the extended function at the starting point.
See Section \ref{sec:hessians} for additional information on
techniques for computing the sparsity
pattern of the extended function.
Once the sparsity pattern is determined,
the compressed AD approach is used on all other iterations.
The hybrid approach is currently the best approach to compute
gradients of partially separable functions, and is used in all solvers
installed on the NEOS Server.

We conclude this section with some recent results on
using the sparse AD approach to compute the gradients of
partially separable functions drawn from the \minpack\ \cite{BMA91a}
collection of test problems.
We selected ten problems; the first five problems are
finite element formulations of variational problems, while the last
five problems are systems of nonlinear equations
derived from collocation or difference formulations of systems
of differential equations.

Table \ref{grads:data} provides the value of $ \rhom{} $ for the
ten problems in our performance results.
For each of the problems we
used three values of $n$,
usually $ n \in \{ 1/4 , 1, 4 \}\cdot 10^ 4 $,
to observe
the trend in performance as the number of variables increases.
The results were essentially independent of the number of
variables, so our results are indicative of the performance
that can be expected in large-scale problems.

\begin{table}[ht]
\begin{center}
\caption{Data for \minpack\ test problems}
\label{grads:data}
\medskip
\begin{tabular}{c|cccccccccc}
 & \textsc{pjb} & \textsc{msa} & \textsc{odc} & \textsc{ssc} & \textsc{gl2} & 
   \textsc{fic} & \textsc{sfd} & \textsc{ier} & \textsc{sfi} & \textsc{fdc} \\
\hline
\rhom{} & 5 & 4 & 4 & 4 & 5 & 9 & 14 & 17 & 5 & 13 \\ 
\end{tabular}
\end{center}
\end{table}

We want to show that the bounds \Ref{psf:bound}
for $ \ctimeg $ holds for these problems.
For these results we used the sparse approach to compute
the Jacobian matrix $ \fext ' (x) $ of the extended function,
and then computed the gradient of $f$ with \Ref{grads:psf}.
For each problem we computed
the ratio $ \kappa_1 $, where
\[
T \left \{ \grad f(x) \right \} = \kappa_1 \,
\rhom{} \max T \{ f(x) \} .
\]
Table \ref{grads:table} presents the quartiles for $ \kappa_1 $ 
obtained on a Pentium 3 (500 MHz clock, 128 MB of memory) 
with the Linux operating system.

\begin{table}[ht]
\begin{center}
\caption{Quartiles for $ \kappa_1 $ on the \minpack\ problems} 
\label{grads:table}
\medskip
\begin{tabular}{ccccc}
$ \min $ & $ q_1 $ & $ q_2 $ & $ q_3 $ & $ \max $ \\
\hline
1.3 & 2.8 & 4.5 & 5.3 & 7.8 \\ \hline
\end{tabular}
\end{center}
\end{table}

\noindent
The results in Table \ref{grads:table} show that the 
bound \Ref{psf:bound} for $ \ctimeg $ holds for 
the \minpack\  problems, with $ \kappa_1 $ small. 

These results are consistent with the results in 
\cite{BOUARICHA94}, where it was shown that 
$ \kappa_1 \in [ 3,15] $ on a SPARC-10 for another
set of test problems drawn from the \minpack\ collection.
Note that in \cite{BOUARICHA94} the ratio $ \kappa_1 $ was
computed with $ \rhom{} $ replaced by the number of columns $p$
in the matrix $V$. Since $ p \ge \rhom{} $, the ratios in
Table \ref{grads:table} would decrease if we replaced $ \rhom{} $ by $p$.
The advantage of using $ \rhom{} $ is that
the ratio $ \kappa_1 $ is then dependent only on the structure
of the function.

\section{Computing Hessian Matrices}

\label{sec:hessians}

We have already shown how automatic differentiation tools 
can be used to compute the gradient of a partially separable function.
We now discuss the tools that are needed
to compute the Hessian of a partially separable so that
the requirements \Ref{psf:ctime} on computing time and
\Ref{psf:cmem} on memory are satisfied.

The techniques that we propose require the
sparsity pattern of the Hessian matrix and that
the Hessian-vector products $ \grad ^ 2 f (x) v $ be available.
In our numerical results we approximate
the Hessian-vector product with a difference of gradient
values, but in future work we expect to compute Hessian-vector
products with ADIFOR.

We now show how to compute the sparsity pattern
of the Hessian matrix from the sparsity pattern of $ \fext' (x) $.
We define the sparsity pattern of a matrix-valued mapping
$ A : \R^n \mapsto \R^{n\times n} $
in a neighborhood $ N ( x_0 ) $ of a point $ x_0 $ by
\begin{equation}
\label{hessians:pattern}
\cS \left \{ A(x_0) \right \} \equiv 
\biggl \{ (i,j): a_{i,j} (x) \not \equiv 0 , \ x \in N(x_0) \biggr \} .
\end{equation}
We are interested in the sparsity
pattern of the extended Jacobian and the Hessian matrix of a partially
separable function $ f : \R^n \mapsto \R $ in
a region $ \cD $ of the form
\[
\cD = \left \{ x \in \R^n : x_l \le x \le x_u \right \} .
\]
Given  $ x \in \cD $, we evaluate the sparsity pattern 
$ \cS \left \{ \fext ' (x) \right \} $
by computing $ \fext ' (\bar x_0 ) $, where 
$ \bar x_0 $ is a random, small 
perturbation of $ x_0 $, for example,
\[
\bar x_0 = ( 1 + \varepsilon ) x_0 + \varepsilon, \qquad 
| \varepsilon | \in [   10^{-6} ,  10^{-4} ] .
\]
Then we can reliably let 
$ \cS \left \{ \fext ' (x) \right \} $ be
the set of $ (i,j) $ such that $ \partial _{i,j} \fext ( \bar x_0) \neq 0 $.
We should not obtain the sparsity pattern of the Jacobian matrix 
by evaluating $ \fext' $ at the starting point $ x_0 $
of the optimization process because this point 
is invariably special, and thus the sparsity pattern of
the Jacobian matrix is unlikely to be representative.

The technique that we have outlined for determining the sparsity
pattern is used by the solvers in the NEOS Server and has proved
to be quite reliable. The sign of $ \varepsilon $ must be
chosen so that $ \bar x_0 \in \cD $, and special care must be taken
to handle the case when $x_l$ and $x_u$ agree in some component.

Given the sparsity pattern of the Jacobian matrix of the
extended function, we determine the sparsity pattern for the
Hessian  $ \grad^2 f(x) $ of the partially separable function $f$ via
\begin{equation}
\label{hessians:sparse}
\cS \left \{ \grad^2 f (x) \right \} \subset 
\cS \left \{ \fext ' (x)^T \fext ' (x)  \right \} .
\end{equation}
Note that \Ref{hessians:sparse} is valid only in terms of
the definition \Ref{hessians:pattern} for a
sparsity pattern. For example,
if $ f : \R^2 \mapsto \R $ is defined by 
\[
f (x) = \phi (\xi_1  \xi_2) 
\]
for a function $ \phi $ such that $ \phi ' (0) \neq 0 $, then
$ \partial _{1,2} f (0) \neq 0 $, but
$ \partial_1 f (0) = \partial_2 f (0) = 0 $.
However, \Ref{hessians:sparse} holds because
$ \partial_2 f (x) \not \equiv 0 $ and
$ \partial_1 f (x) \not \equiv 0 $ in a neighborhood
of the origin.

In most cases equality holds in \Ref{hessians:sparse}.
This happens, in particular,
if $f$ does not depend linearly on the
variables, and
\begin{equation}
\label{hessians:nocancel}
\bigcup _{k=1}^{m} \cS \left \{ \grad^2 f_k (x) \right \}  \subset \cS 
\left \{ \grad^2 f (x) \right \} .
\end{equation}
If $ f $ depends linearly on some variables, say,
\[
f(x) =  \xi_1  + \phi ( \xi_2 , \ldots , \xi_n ) ,
\]
then equality does not hold in \Ref{hessians:sparse}.
Assumption \Ref{hessians:nocancel} implies that there
is no cancellation in the computation of the
Hessian $ \grad^2 f (x) $. This assumption can fail in some
cases, for example, when $ f_1 \equiv - f_2 $, but holds in most cases.

Since we are able to estimate the sparsity pattern of the
Hessian matrix via \Ref{hessians:sparse}, we could use the
compressed AD approach described in Section \ref{sec:gradients}
to compute the Hessian matrix from a compressed Hessian
$ \grad ^ 2 f(x) V $.
However, these techniques ignore the symmetry of the Hessian
matrix and thus may require an unnecessarily large number of
columns $p$ in the matrix $V$.
For example, an arrowhead matrix requires $ p = n $ if symmetry
is ignored, but $ p = 2 $ otherwise.

Powell and Toint \cite{MJDP79} were the first to 
show that symmetry can be used to reduce the number $p$ of
columns in the matrix $V$. They proposed two methods
for determining a symmetric matrix $A$ from a compressed
matrix $AV$.
In the direct method the unknowns in $A$ are determined directly
from the elements in the compressed matrix $ AV $. In this
method unknowns are determined independently of each other.
In the substitution method the unknowns are determined in a given
order, either directly or as a linear combination of elements
that have been previously determined.

These definitions of direct and substitution methods
are precise but do not readily yield algorithms for
determining symmetric matrices.
Coleman and Mor\'e \cite{TFC84b} and Coleman and Cai \cite{TFC86}
extended \cite{MJDP79} by interpreting
the problem of determining symmetric matrices
in terms of special graph coloring
problems. This work led to new algorithms and a deeper
understanding of the estimation problem.

Software for the symmetric graph coloring problem is
available \cite{TFC85} for both direct and substitution methods.
Numerical results in \cite{TFC84b} suggest that a direct
method yields a 20\% improvement over methods
that disregard symmetry,
and that the substitution method yields about
a 30\% reduction over the direct method.

We use Algorithm \ref{hessians:adhess} 
to compute the Hessian matrix
from a user-supplied extended function $ \fext $.
This algorithm uses static memory allocation so that
it is first necessary to determine the number of nonzeros in
$ \fext ' (x_0) $ by
computing $ \fext ' (x_0) $ by rows, but not storing the entries.
Once this is done, we allocate space for $ \fext ' (x_0) $
and compute $ \fext ' (x_0) $ and the sparsity pattern.
Another interesting aspect of 
Algorithm \ref{hessians:adhess} is that we compute
the number of nonzeros in $ \fext ' (x_0)^T \fext ' (x_0) $
directly from the sparsity pattern of $ \fext ' (x_0) $.
In view of \Ref{hessians:sparse}, we then have an accurate
idea of the amount of memory needed to store the Hessian matrix.
The final step is to compute the Hessian matrix from the
the compressed Hessian matrix
$ \grad^2 f (x_0 )V $ by either a direct or a substitution method.

\begin{Algorithm}
\begin{list}{$\diamond$}{\setlength{\itemsep}{0pt}}
\item
Evaluate $ \fext (x_0) $ and obtain $ m = \mbox{size} \fext(x_0) $. 
\item
Compute  $ \mbox{nnz} \{ \fext ' (x_0) \} $.
\item
Allocate space for $ \fext ' (x_0) $.
\item
Compute the sparsity pattern $ \cS \{ \fext ' (x_0) \} $.
\item
Compute $ \mbox{nnz} \{ \fext ' (x_0)^T \fext ' (x_0) \} $.
\item
Allocate space for $ \grad^2 f (x_0 )$
\item
Compute $ \grad^2 f (x_0 )$ from the compressed Hessian matrix
$ \grad^2 f (x_0 )V $.
\end{list}
\caption{Computing the Hessian matrix for a partially separable function.}
\label{hessians:adhess}
\end{Algorithm}

We consider both direct and substitution methods to determine
the Hessian matrix from the compressed Hessian.
In both cases we are interested in the
ratio $ \kappa_2 $, where
\[
T \left \{ \grad ^ 2 f(x) \right \} = \kappa_2  
 \rhom{^2}  T \{ f(x) \} ,
\]
since this provides a measure of the cost of evaluating
the Hessian matrix relative to the cost of the function.
The $ \kappa_2 $ quartiles for both direct and 
substitution methods  on the \minpack\ problems
used in Section \ref{sec:gradients}
appear in Table \ref{hessians:table}.

\begin{table}[ht]
\caption{Quartiles for $ \kappa_2 $ on 
{\sf MINPACK-2} problems}
\label{hessians:table}
\begin{center}
\begin{tabular}{rccccc}
Method & $ \min $ & $ q_1 $ & $ q_2 $ & $ q_3 $ & $ \max $ \\
\hline
Direct       & 1.6 & 5.1 & 11.2 & 15.2 & 46.4 \\ \hline 
Subsitution  & 1.5 & 4.1 & 9.0 & 12.5 & 30.2 \\ \hline
\end{tabular}
\end{center}
\end{table}

Direct and substitution methods usually require
more than $ \rhom{} $ gradient evaluations to determine
the Hessian matrix, and thus the increase in the value of 
$ \kappa_2 $ relative to $ \kappa_1 $ in Table \ref{grads:table}
was expected.
Still, it is reassuring that the median value of $ \kappa_2 $ is reasonably small.
The largest values of $ \kappa_2 $ are due to one of the problems; if
this problem is eliminated, then the maximal value drops by at least a factor of two.
In general, problems with the
longest computing times yield the smallest values of $ \kappa_2 $
since these problems tend to mask the overhead in the automatic
differentiation tools and in determining the Hessian matrix.
Moreover, these results are based on using
a gradient evaluation that relies on sparse
automatic differentiation tools; the use of the
hybrid approach mentioned in Section~\ref{sec:gradients}
should reduce $ \kappa_2 $ substantially.

\section*{Acknowledgments}

\label{acks}

Paul Hovland merits special mention for sharing 
his considerable knowledge of automatic differentiation tools.
Liz Dolan used a preliminary implementation of the techniques
in this paper to install \tron\ on the NEOS Server, and in the
process sharpened these techniques.
Gail Pieper provided the final touches on the paper with her
careful editing.

\bibliographystyle{siam}

\bibliography{ad2k}

\end{document}